**Optical bulk-boundary dichotomy in a quantum spin Hall insulator**

Junfeng Han[1,2,3†], Pengcheng Mao[1,4†], Hailong Chen[5,6,7†], Jia-Xin Yin[8†], Maoyuan Wang[1,9†], Dongyun Chen[1,3], Yongkai Li[1,2,3], Jingchuan Zheng[1,2,3], Xu Zhang[1,2,3], Dashuai Ma[1,10], Qiong Ma[11], Zhi-Ming Yu[1,2,3], Jinjian Zhou[1,2,3], Cheng-cheng Liu[1,2,3], Yeliang Wang[12], Shuang Jia[13], Yuxiang Weng[5,6], M. Zahid Hasan[8], Wende Xiao[1,2,3]*, Yugui Yao[1,2,3]*

[1]Key laboratory of advanced optoelectronic quantum architecture and measurement, ministry of education, School of Physics, Beijing Institute of Technology, Beijing, 100081, China.

[2]Yangtze Delta Region Academy of Beijing Institute of Technology, Jiaxing, 314000, China.

[3]Beijing Key Lab of Nanophotonics & Ultrafine Optoelectronic Systems, Beijing Institute of Technology, Beijing, 100081, China.

[4]Analysis & Testing Center, Beijing Institute of Technology, Beijing 100081, China.

[5]Beijing National Laboratory for Condensed Matter Physics and Institute of Physics, Chinese Academy of Sciences, Beijing, 100190, China.

[6]Songshan Lake Materials Laboratory, Dongguan, Guangdong, 523808, China.

[7]School of Physical Sciences, University of Chinese Academy of Sciences, Beijing, 100190, China.

[8]Laboratory for Topological Quantum Matter and Advanced Spectroscopy (B7), Department of Physics, Princeton University, Princeton, New Jersey, USA.

[9]Department of Physics, Xiamen University, Xiamen 361005, China

[10]Department of Physics, Chongqing University, Chongqing 400044, China

[11]Department of Physics, Boston College, Chestnut Hill, MA, USA.

[12]School of Integrated Circuits and Electronics, MIIT Key Laboratory for Low-Dimensional Quantum Structure and Devices, Beijing Institute of Technology, Beijing 100081, China.

[13]ICQE, School of Physics, Peking University, Beijing, 100871, China.

**Abstract:**

The bulk-boundary correspondence is a key concept in topological quantum materials. For instance, a quantum spin Hall insulator features a bulk insulating gap with gapless helical boundary states protected by the underlying $Z_2$ topology. However, the bulk-boundary dichotomy and distinction are rarely explored in optical experiments, which can provide unique information about topological charge carriers beyond transport and electronic spectroscopy techniques. Here, we utilize mid-infrared absorption micro-spectroscopy and pump-probe micro-spectroscopy to elucidate the bulk-boundary optical responses of $Bi_4Br_4$, a recently discovered room-temperature quantum spin Hall insulator. Benefiting from the low energy of infrared photons and the high spatial resolution, we unambiguously resolve a strong absorption from the boundary states while the bulk absorption is suppressed by its insulating gap. Moreover, the boundary absorption exhibits a strong polarization anisotropy, consistent with the one-dimensional nature of the topological boundary states. Our infrared pump-probe microscopy further measures a substantially increased carrier lifetime for the boundary states, which reaches one nanosecond scale. The nanosecond lifetime is about one to two orders longer than that of most topological materials and can be attributed to the linear dispersion nature of the helical boundary states. Our findings demonstrate the optical bulk-boundary dichotomy in a topological material and provide a proof-of-principal methodology for studying topological optoelectronics.

†These authors contributed equally to this work.

*Email: wdxiao@bit.edu.cn, ygyao@bit.edu.cn.

## 1. Introduction

The quantum spin Hall insulators have one-dimensional (1D) dissipationless conducting channels in the edge due to topological protection from backscattering, which is widely expected to be a promising platform for next-generation optoelectronics with high speed and high efficiency [1-10]. However, the critical scientific issue of carrier dynamics of 1D edge states, which inevitably involves dynamic controls of nontrivial carriers, has rarely been addressed. Though 1D edge states can be probed by transport and electronic spectroscopy techniques [11-20], however, most optical and photoelectric methods have no such high spatial resolution to distinguish the properties of bulks and boundaries, which hinders both research and application of 1D topological edge states in the field of optoelectronics. One strategy to solve this problem is to stack quantum spin Hall insulator layers along the direction normal to the layer plane, as the existence of multiple 1D topological edge states can enhance the response signals. Meanwhile, this method also requires a weak coupling between the edge states localized at each layer.

Bismuth halogenides, composed of 1D chains (Fig. 1**a**), were predicted as topological materials with various fascinating properties [21-36]. In particular, the single-layer $Bi_4Br_4$ was identified as a large gap quantum spin Hall insulator that hosts gapless helical edge states [21-24]. As the interlayer coupling of $Bi_4Br_4$ is very weak, the topological characters of the edge states in multiple layers are essentially preserved [24]. Thus, the bulk $Bi_4Br_4$ with multiple edge states can enhance the response signals of 1D helical edge states. For instance, the topological boundary states in rough $Bi_4Br_4$ samples with abundant edges and hinges were probed by angle-resolved photoemission spectroscopy [33], a widely adopted technique with rather low spatial resolutions (ca. 1 μm). As the spot sizes of ultraviolet and infrared lights can be tuned to the same order, it is technically possible to probe the 1D edge states by infrared techniques. Here, we utilize mid-infrared absorption micro-spectroscopy to elucidate the real-space optical response of $Bi_4Br_4$. Moreover, the pump-probe micro-spectroscopy allows us to experimentally observe the carrier dynamic process of topological edge states compared with bulk states. Our work constitutes the first report on the topological edge states by optical methods, paving the way for the application of similar techniques.

## 2. Materials and methods

Single crystals of $Bi_4Br_4$ were synthesized by the Bi self-flux method. The crystals up to $2\times0.5\times0.2$ mm$^3$ in size were obtained in the sealed ampoule. XRD measurements were used to characterize the crystal structure of $Bi_4Br_4$. The surface morphology was characterized by AFM (Bruker Multimode 8). The chemical composition of the $Bi_4Br_4$ belts was analyzed by SEM with EDS (Zeiss Gemini 550). The chemical composition of the Bi4Br4 belt was also examined by XPS (FEI QUANTERA SXM). The optical image and infrared absorption spectra were acquired using a Fourier transform infrared microscopy (FTIR, HYPERION 3000 Microscopy, Bruker) with a pair of 15× objectives (the detection spectral range from 4000 $cm^{-1}$ to 900 $cm^{-1}$). The carrier dynamics in the energy range around the band gap of $Bi_4Br_4$ was investigated by ultrafast infrared pump-probe micro-spectroscopy was performed with a femtosecond Ti: Sapphire regenerative amplifier (Spitfire Ace, 3.5W, Spectra Physics). Detailed experimental measurements are described in the supplementary materials (online). All optical measurements are carried out at room temperature.

## 3. Results and discussion

We first use a Fourier transform infrared microscopy to detect the infrared spectra of $Bi_4Br_4$ ribbons with photon energy ranges from 0.50 eV to 0.11 eV at room temperature (Fig. 1**b**). The $Bi_4Br_4$ ribbon can be easily obtained via mechanical exfoliation from a single crystal [37]. In the upper panel of Fig. 1**c**, a typical ribbon can be seen in optical microscope images with 0.8 μm in thickness, 40 μm in wideness, and more than 100 μm in length. A spatial resolution of the infrared absorption maps is estimated to be 7-8 μm. In the lower panels of Fig. 1**c**, we first obtain an absorption micro-spectroscopy mapping image with a photon energy of 0.24 eV and observe uniform absorption across the entire flake (more precisely, gradually decay from bulk to edge because the laser is moving away from the sample). In sharp contrast, the same absorption mapping image obtained with a photon energy of 0.20 eV shows strong absorption only close to edges with the bulk signals being significantly suppressed. This observation directly demonstrates that edge and bulk have distinct dependence on photon energy.

In order to systematically understand the photon energy dependence, we respectively focus the laser at the bulk and edge, and measure their absorption as a function of photon energy between 0.12 eV and 0.48 eV. In Fig. 1**d**, the absorption of the bulk exhibits a dramatic reduction when the photon energy is smaller than 0.24 eV while the absorption of the edge remains relatively flat. This reduction of the bulk absorption can be attributed to the bandgap of bulk electronic states of $Bi_4Br_4$. We estimate a bandgap of 0.22 eV by using a classic Tauc plot of $(\alpha h\nu)^2$ vs. $h\nu$ (Fig. S7 on line) [38]. The optical band gap is in the range of the reported theoretical and experimental results. The calculated energy gap of bulk electronic structure is around 0.145-0.3 eV [30,33], while the tunneling differential conductance reveals an insulating gap of 0.26 eV [23] and the angle-resolved photoemission spectroscopy measurements on the cleaved surface of $Bi_4Br_4$ shows a bandgap of 0.23-0.3 eV [33,39]. The absorption of the edge persists to very low photon energies. Such robust edge absorption can be attributed to the gapless topological boundary states.

Moreover, due to the anisotropic structure of $Bi_4Br_4$, the anisotropic optical transition matrix elements give rise to the anisotropic infrared absorption. To shed light on the anisotropic infrared absorption of $Bi_4Br_4$, we take an absorption map by using linearly polarized light nominally perpendicular to the top (*ab*-plane) of the $Bi_4Br_4$ belt with the polarization direction along the *a*- or *b*-axis with the photon energies of 0.20 eV and 0.24 eV (Fig. 2 **a** and **b**). The absorbance decreases with the polarization varying from the $\mathbf{E} \parallel \mathbf{b}$ direction to the $\mathbf{E} \parallel \mathbf{a}$ direction, especially at the boundaries with the photon energy of 0.20 eV. In Fig. 2**c**, theoretical calculations of optical absorption of $Bi_4Br_4$ belt indicate a strong anisotropic infrared absorption both at the belt boundary and bulk, in reasonable agreement with our experimental observation. Furthermore, we defined absorption anisotropy as $(\alpha_{E//b}-\alpha_{E//a})/(\alpha_{E//b}+\alpha_{E//a})$ and compare those values from belt boundary and bulk, respectively. As shown in Fig. 2**d**, the boundary has stronger absorption anisotropy than that in the bulk, especially with photon energy smaller than the theoretical bandgap.

After identifying the difference between bulk and boundary optical responses with diffraction-limited infrared microscopy, we further use infrared pump-probe techniques to explore their respective carrier dynamics. Figures. 3 **a** and **b** illustrate our infrared pump-probe micro-spectroscopy setup. The key techniques are to obtain a wide-energy-range probe light by forming a plasma filament (0.12-0.45 eV) and to detect the wide spectra using mercury-cadmium-telluride arrays. This powerful technique allows us to inspect the carrier dynamics of the bulk and boundary of the $Bi_4Br_4$ belt in multi-dimensional parameter space (20 μm spatial resolution, wide photon energy range of 0.12-0.45 eV, and wide time range of 0.1 ps - 3000 ps).

In Fig. 3**c**, the pump photon energy is fixed at 0.17 eV to avoid probing interband transitions between the bulk states. The evolution of the photo-excited carriers is then monitored by recording the transmittance variation ($\Delta T/T$). It is clear that the negative signals are significant around the boundary and decayed quite slowly with the probe photon energy of 0.15-0.18 eV. Moreover, these negative signals can still be clearly observed after 1.5 ns relaxation, manifesting as an ultralong carrier lifetime. In Fig. 3**d**, the positive signals appear in the bulk and decay quickly in tens of picoseconds with the probe photon energy of 0.15-0.18 eV. Those results clearly demonstrate the different dynamics of the bulk and boundary. Furthermore, the typical decay curves of $\Delta T/T$ at the bulk and the boundary are collected and fitted with the multiple-exponential functions (Fig. 3**e**). In the bulk, the positive signal of $\Delta T/T$ pumped by 0.5 eV photons decays in ps timescale. The relaxation is decomposed into two components: a fast one ($\tau_1$ = 3~5 ps) and a slow one ($\tau_2$ = 50~100 ps). In contrast to the bulk, the signal from the belt boundary has negative $\Delta T/T$ with an ultralong component ($\tau_3$ > 1 ns) of carrier lifetime. To understand the distinguished behavior of bulk and boundary, a schematic image of carrier excitation is shown in Fig. 3**f**. First, the probe photons with energy smaller than the bandgap can only be absorbed by the carriers in the boundary states or the free carriers in the conduction band. In the belt boundary, the decreased carrier density in the boundary states reduces the infrared absorption after optical excitation and consequently leads to negative signals. This understanding is also consistent with the observation of strong infrared absorption at the belt boundaries with photon energy < 0.22 eV (the bandgap of $Bi_4Br_4$). In the bulk, more photo-excited carriers in the conduction band can enhance the absorption of probe photons, contributing to the positive signals.

We systematically repeat the measurements at the boundary and the bulk with the pulse pump fluences varying from 2.1 μJ/cm$^2$ to 304 μJ/cm$^2$ (Fig. 4**a**). By fitting those decay curves with the multiple-exponential functions, we can extract the delay times of both at the boundary and the bulk, which have a strong dependence on the pump fluences (Fig. 4**b**). The faster decay is observed with stronger pump fluences, which is caused by the enhanced electron-electron scattering or auger recombination with more hot-carriers [40,41]. The longest carrier lifetime of 1.5 ns at room temperature is observed in the boundary with low pump fluences of 2.1 $\mu J/cm^2$. The nanosecond lifetime is about one to two orders longer than that of the bulk and most topological materials [40-47]. In Fig. 4**c**, the magnitude of the signal from the boundary is performed with a saturation behavior, while the magnitude of the signal in bulk is proportional to the pump fluences. A fundamental model can be used to describe the pump fluences dependent signals: $\Delta T/T \propto E/(E+E_s)$, where E and $E_s$ are the pump and saturation fluences, respectively [48]. The fitted curve indicates saturation fluences of ~35 $\mu J/cm^2$. Due to the limited density of edge states, the photons in the incident light are sufficient to excite most carriers of the edge states, leading to the saturation.

The relaxation dynamics of the photoexcited carriers in the $Bi_4Br_4$ belt are discussed in Fig. 4**d**. In the boundary states, the excited carriers in the helical states may relax via intralayer or interlayer scattering. In the intralayer scattering process, the backscattering by the non-magnetic disorder and electron-phonon interaction and the direct recombination of the electrons and holes requiring spin flips, which, however, is incompatible with the helical nature of the boundary states and thus strongly suppressed [49-51]. Therefore, the excited carriers are more likely to be relaxed within one of the two branches via scattering with phonons. This scattering process suffers from a very limited scattering phase space (phonon energy < 20 meV, very few final states available within the one-dimensional channels), leading to an ultralong relaxation time. In the interlayer scattering process, the edge states localized at different layers are weakly coupled or even decoupled in case the edges of the adjacent layers are not strictly aligned [21,24]. The hinge states localized at different hinges are also decoupled. Thus, both the direct recombination of the electrons and holes at different boundaries and the interlayer scattering are suppressed. Therefore, the ultralong carrier lifetimes is more likely to arise from the helical nature of the one-dimensional topological boundary states of the $Bi_4Br_4$ belt. Another possible mechanism is based on the model of Luttinger liquid. When the photon excitation induce the conversion of edge mode into Luttinger liquid due to its one-dimensional feature, the excited states can form a robust and stable states for a long time. Then, considering the weak coupling between edge states with bulk states, those stable states can also be relaxed slowly by scattering with bulk states. In addition, at the belt center where the topological edge states are absent, the relaxation processes are similar to the typical carrier relaxation behaviors commonly observed in semiconductor materials [52]: Process one ($\tau_1$: 3 ~ 5 ps) is the excited electrons relax to band edge and process two ($\tau_2$: 50 ~ 100 ps) is the relaxed electrons recombine with holes in the valence band (more details in supplementary Fig. 10a).

## 4. Conclusions

To conclude, we utilize mid-infrared absorption micro-spectroscopy and pump-probe micro-spectroscopy to elucidate the bulk-boundary optical responses of quantum spin Hall insulator $Bi_4Br_4$. A strong absorption is resolved from the boundary states while the bulk absorption is suppressed by its insulating gap. Moreover, the boundary absorption exhibits a strong polarization anisotropy, consistent with the one-dimensional nature of the topological boundary states. The infrared pump-probe microscopy further measures a substantially increased carrier lifetime for the boundary states, which is about two orders longer than that of the bulk. Our work taken together demonstrates the optical bulk-boundary dichotomy in a topological material $Bi_4Br_4$.

**Conflict of interest**

The authors declare that they have no conflict of interest.

**Acknowledgments**

This work is funded by the National Natural Science Foundation of China (11734003, 62275016, 12274029 and 92163206), the National Key Research and Development Program of China (2020YFA0308800), Beijing Natural Science Foundation (Z210006, Z190006) and the Strategic Priority Research Program of Chinese Academy of Sciences (XDB30000000). We are grateful to the Instrument Analysis Center of Xi'an Jiaotong University for assistance with infrared absorption measurement and Analysis & Testing Center of Beijing Institute of Technology for assistance with SEM and XRD analyses. We acknowledge fruitful discussions with Xiang Li, Junxi Duan, Qingsheng Wang, Gang Wang, Jie Ma and Yuanchang Li.

**Author contributions**

Y.G.Y., J.F.H. and W.D.X. supervised this project; P.C.M., H.L.C. and J.F.H. carried out infrared absorption measurements; D.Y.C., Y.K.L., J.C.Z., X.Z., Y.L.W., S.J., Y.X.W. and W.D.X. synthesized and characterized materials; M.Y.W., D.S.M., Z.M.Y., J.J.Z., C.C.L. and Y.G.Y. performed first-principles calculations; J.F.H., J.X.Y., Q.M., M.Z.H., M.Y.W., W.D.X. and Y.G.Y. wrote the paper; all authors discussed and analyzed the data.

**Figures and figure captions:**

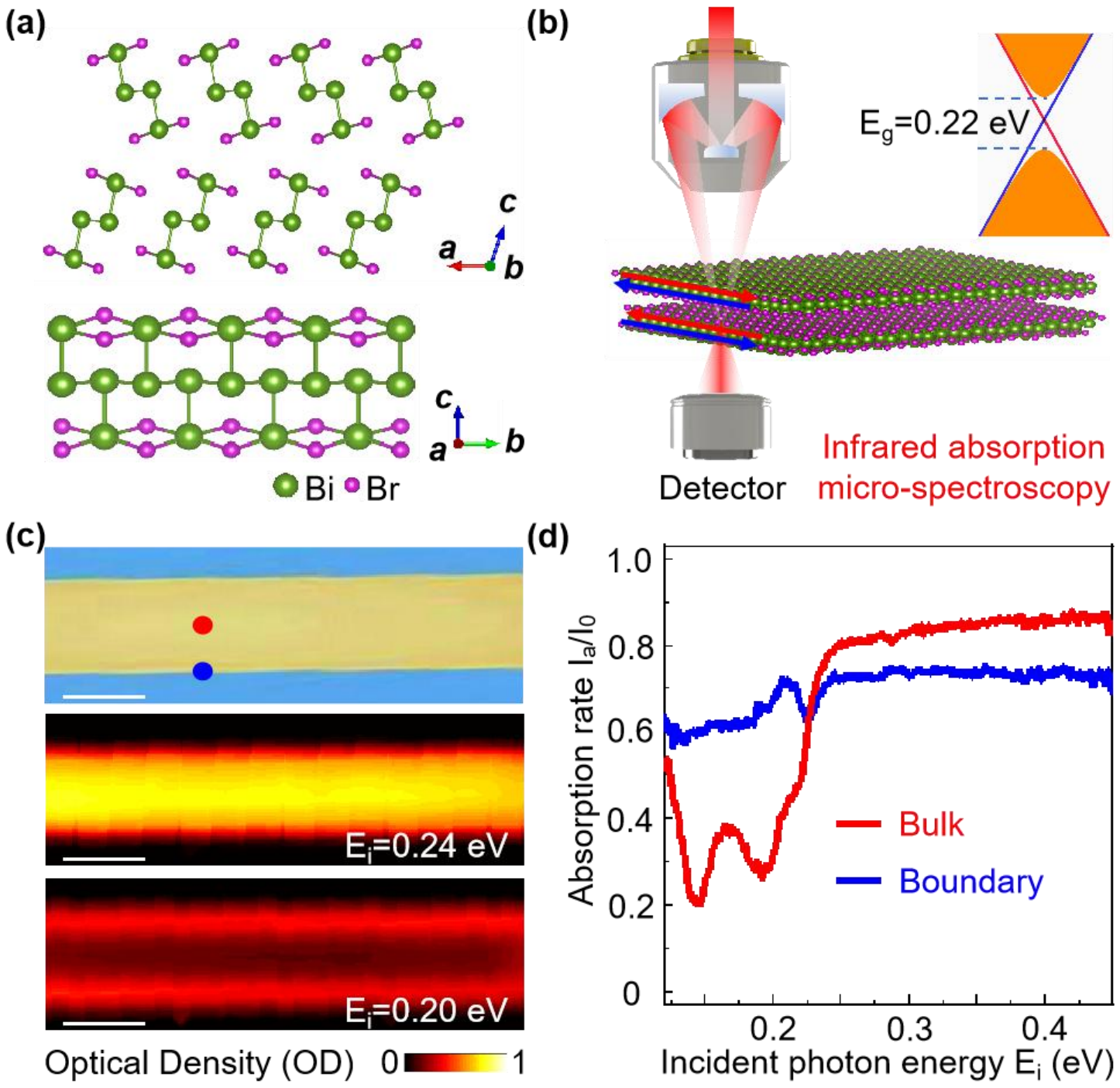

**Fig. 1.** Bulk-boundary infrared absorption. (a) The crystal structure of $Bi_4Br_4$ from two side views. (b) Schematic of infrared absorption micro-spectroscopy. The incident photon energy range is from 0.12 eV to 0.48 eV. The optical band gap of $Bi_4Br_4$ is around 0.22 eV. (c) Infrared absorption micro-spectroscopy mapping. The upper panel shows the optical image of the $Bi_4Br_4$ belt. The lower two panels show the optical density (OD) maps with the incident photon energies $E_i$ = 0.24 eV and 0.20 eV, which is respectively higher and lower than the bandgap of $Bi_4Br_4$. OD is defined as $\log(I_0/I)$, where $I_0$ and I are the intensities of the incident and transmission light, respectively. The Strong infrared absorption of 0.20 eV photons at the boundary can be naturally accounted for by the gapless topological boundary states, which reveals the bulk-boundary dichotomy in $Bi_4Br_4$. Scale bar: 40 μm. (d) Infrared absorption spectra acquired from the bulk (red lines) and the boundary (blue lines). The absorption of the bulk exhibits a dramatic reduction when the photon energy is smaller than 0.24 eV while the absorption of the edge remains flat.

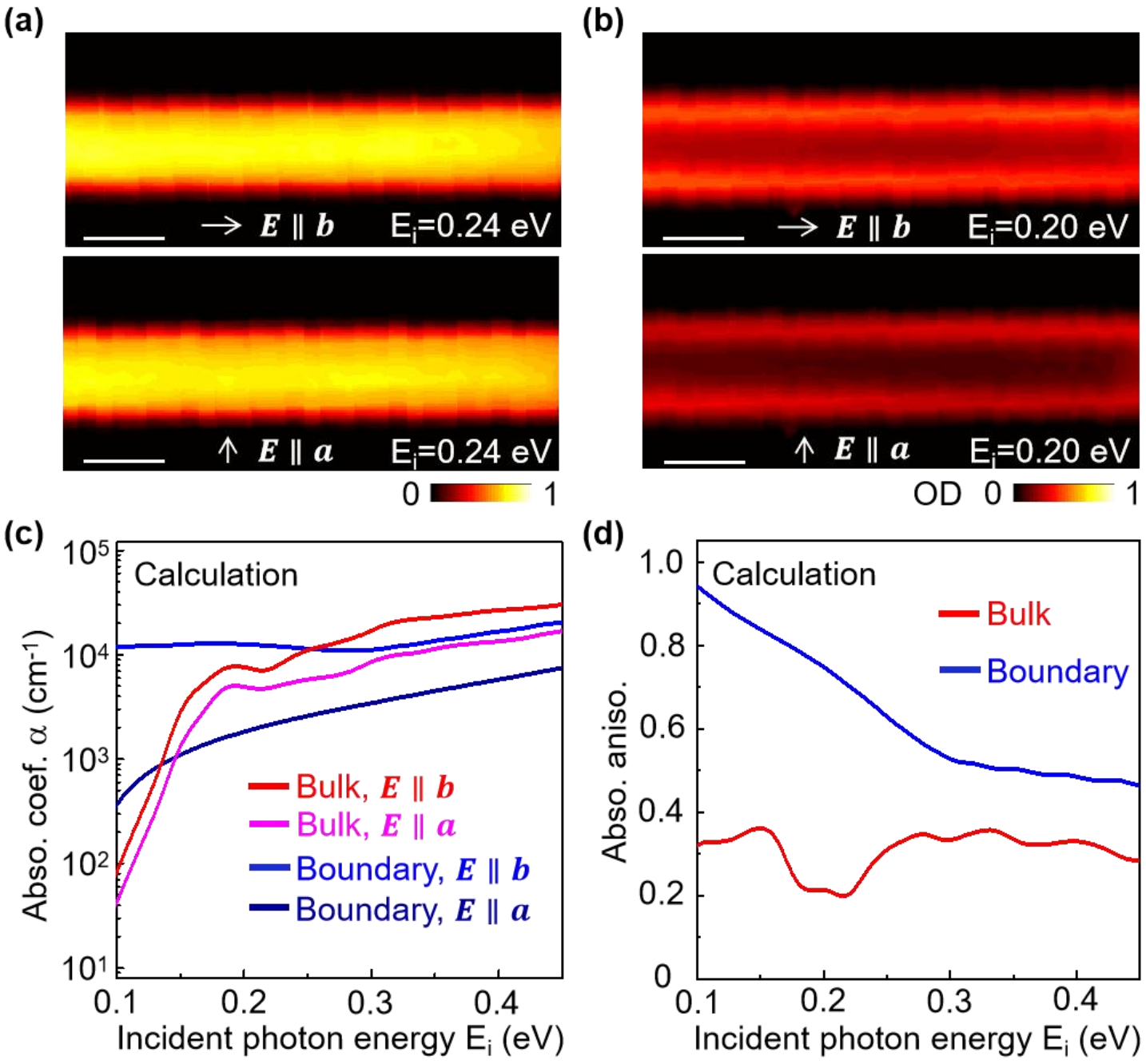

**Fig. 2.** Bulk-boundary anisotropy absorption of $Bi_4Br_4$. (a,b) Optical density maps taken with different polarized incident lights *E*//*b* and *E*//*a*, respectively, where their intensity difference is consistent with the expected bulk and boundary absorption anisotropy in theory. Each scale bar denotes 40 μm. (c) Calculated bulk (red and pink line) and boundary (blue and Royal line) infrared absorption coefficient α as a function of the incident light energy with *a*- and *b*-polarized incident light. The theoretical bandgap of $Bi_4Br_4$ is 0.145 eV by using HSE06. Both bulk and boundary show a strong anisotropic infrared absorption. (d) Calculated bulk (red line) and boundary (blue line) anisotropy infrared absorption as a function of the incident light energy. The absorption anisotropy is defined as $(\alpha_{E//b}-\alpha_{E//a})/(\alpha_{E//b}+\alpha_{E//a})$. The boundaries have stronger absorption anisotropy than that in the bulk, especially with photon energy smaller than the bandgap of 0.145 eV.

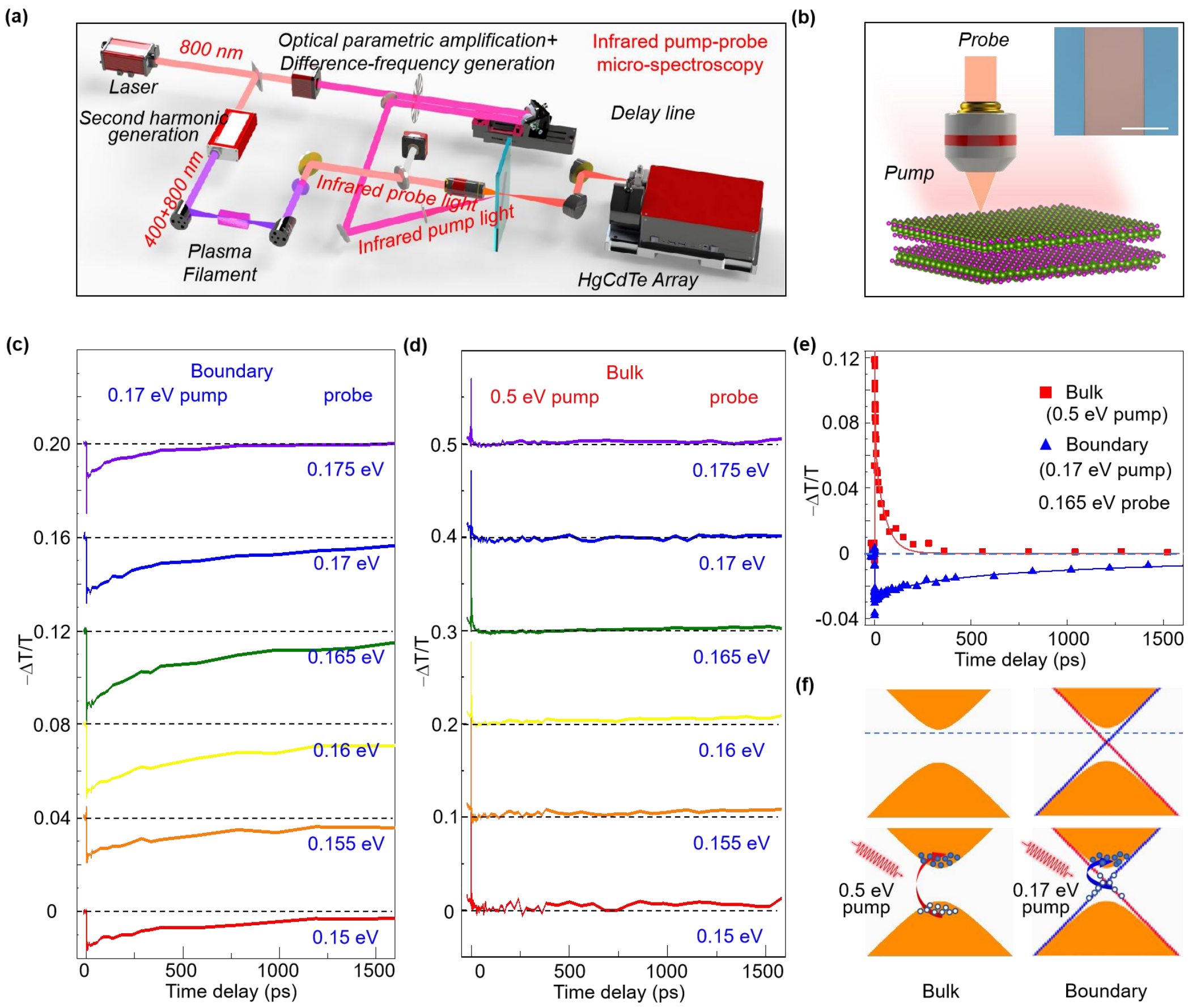

**Fig. 3.** Infrared pump-probe micro-spectroscopy of bulk-boundary states. (a) Illustration of our infrared pump-probe micro-spectroscopy setup. The 35 fs pulses at a repetition rate of 1 kHz were generated by femtosecond amplifier, split and transformed to the wide-range infrared pump and probe light, focused and transmitted successively through the sample, and detected by an infrared array detector. (b) Schematic of detection process of infrared pump-probe micro-spectroscopy. The inset shows the optical image of the crystal. The scale bar denotes 40 μm. The thickness of the belt is 1.3 μm. (c,d) Temporal evolutions of the 0.5 eV and 0.17 eV excitation-induced ΔT/T detected with $E_{probe}$ = 0.15 ~ 0.18 eV at the $Bi_4Br_4$ boundary and bulk. The negative signals are significant around the boundary and decay quite slowly, while the positive signals appear in the bulk with quickly decaying in tens of picoseconds. (e) Typical temporal evolutions of 0.5 eV and 0.17 eV excitation-induced ΔT/T at the boundary and bulk. The raw data (dots) are multi-exponentially fitted (curves). An obvious ultralong negative signal can be observed at the boundary. (f) Schematic of excitation process by 0.5 eV and 0.17 eV photons at the bulk and boundary, respectively. The excited bulk state has more free carriers, which can enhance the absorption of 0.165 eV photons to form positive signals. The excited boundary state has fewer carriers, which reduces the infrared absorption and consequently leads to negative signals.

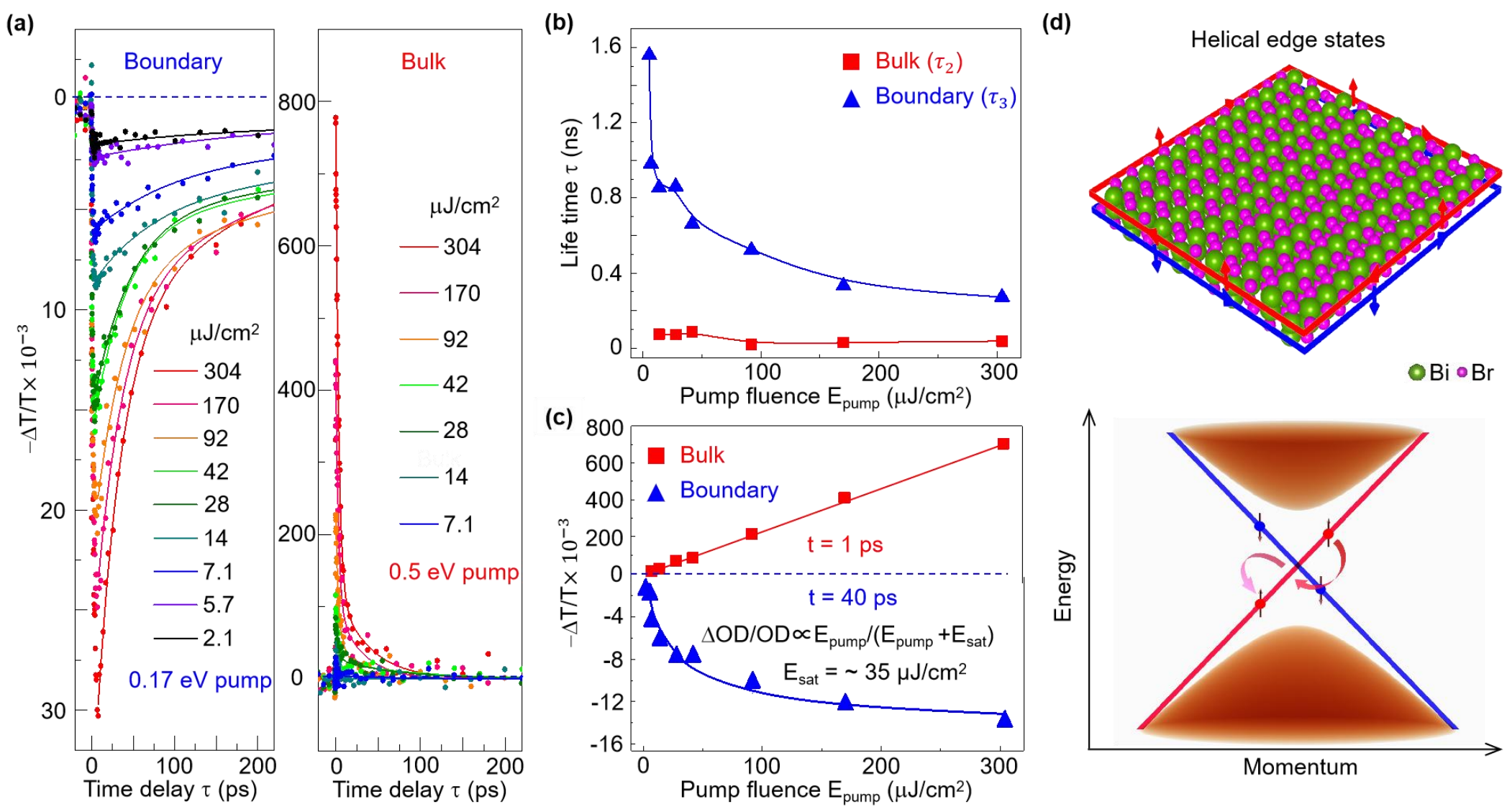

**Fig. 4.** Bulk-boundary lifetimes. (a) Temporal evolutions of transmittance change (ΔT/T) pumped with pump energy $E_{pump}$ = 2 - 304 μJ/cm$^2$ of each laser pulse at the boundary and bulk. (b) The carrier lifetime τ as a function of the pump energies. The boundary carrier lifetime at low pump fluences reaches one nanosecond scale at room temperature. (c) The magnitude of the ΔT/T as a function of $E_{pump}$. The blue solid line indicates a saturation function fit for the signals from the belt boundary ($\tau_3$), the saturation energy $E_{sat}$ = 35 μJ/cm$^2$. The red line indicates a linear relationship between ΔT/T and $E_{pump}$ in the bulk ($\tau_2$). (d) Schematic of helical edge states and the schematic representation of relaxation processes of photoexcited carriers in the boundary states. The electron-phonon scattering with the same spin polarization is suppressed by the very limited scattering phase space, leading to an ultralong carrier lifetime at the boundary.

## Electronic Supplementary Materials

### Supplementary Note 1: Preparation and characterization of α-$Bi_4Br_4$ belts

Single crystals of α-$Bi_4Br_4$ were grown by a Bi self-flux method. The crystals up to 2×0.5×0.2 $mm^3$ in size were obtained in sealed ampoules. Supplementary Fig. 1 shows an optical image of the as-grown single crystals. Their needle shapes indicate a strong anisotropy of the crystal structure. X-ray diffraction (XRD) confirm that the as-grown crystals are α-$Bi_4Br_4$, as shown in supplementary Fig. 2. The chemical compositions of α-$Bi_4Br_4$ crystals were analyzed with a field emission scanning electron microscope (SEM, Zeiss Gemini 550) with an energy-dispersive spectrometer (EDS, Supplementary Fig. 3) and X-ray photoelectron spectroscopy (XPS, FEI QUANTERA-II SXM, Supplementary Fig. 4). A Bi:Br atomic ratio of 1:1 was quantitied by EDS analysis. Quantitative analysis based on the XPS spectra of the Bi 4f and Br 3d core levels of α-$Bi_4Br_4$ crystals also revealed a Bi/Br atomic ratio of 1:1. The thickness and step profiles were determined using a profilometer (Bruker Dektak XT). The surface morphology was characterized by AFM (Bruker Multimode 8).

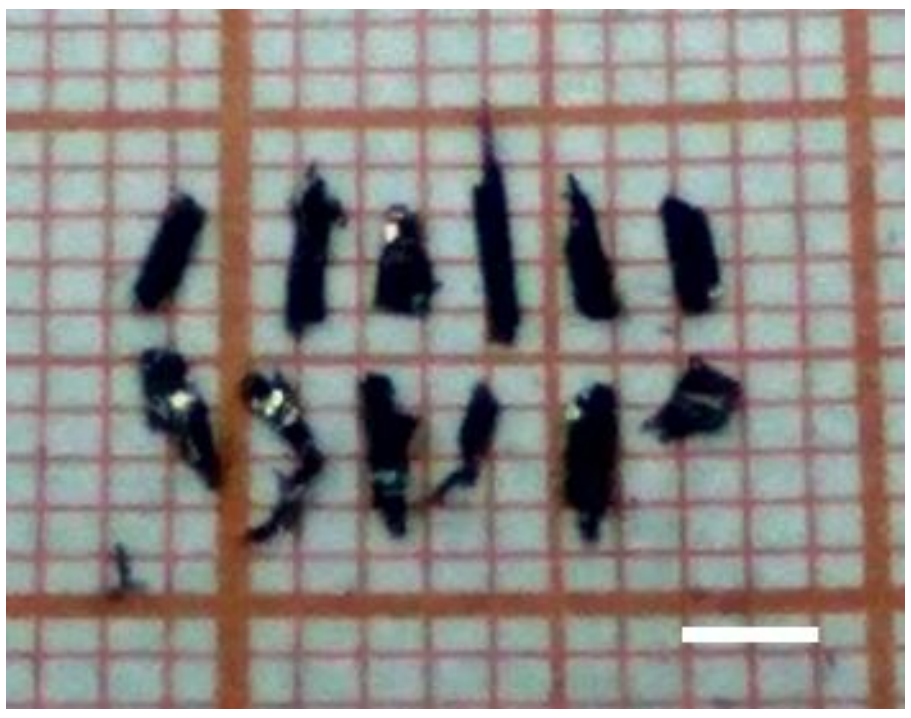

**Supplementary Figure 1 |** Optical image of the as-grown α-$Bi_4Br_4$ single crystals. Scale bar, 2 mm.

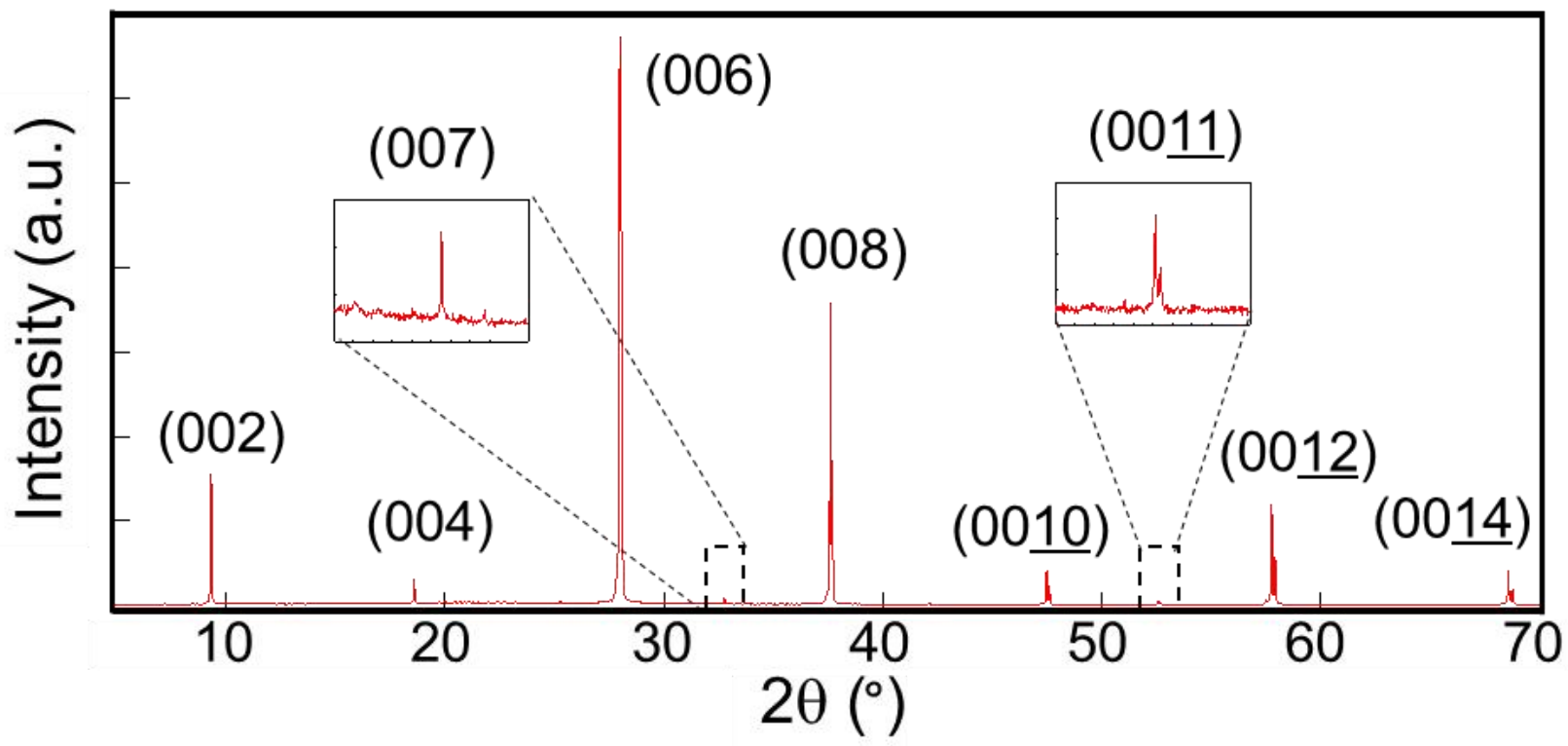

**Supplementary Figure 2 |** XRD pattern of α-$Bi_4Br_4$ crystal.

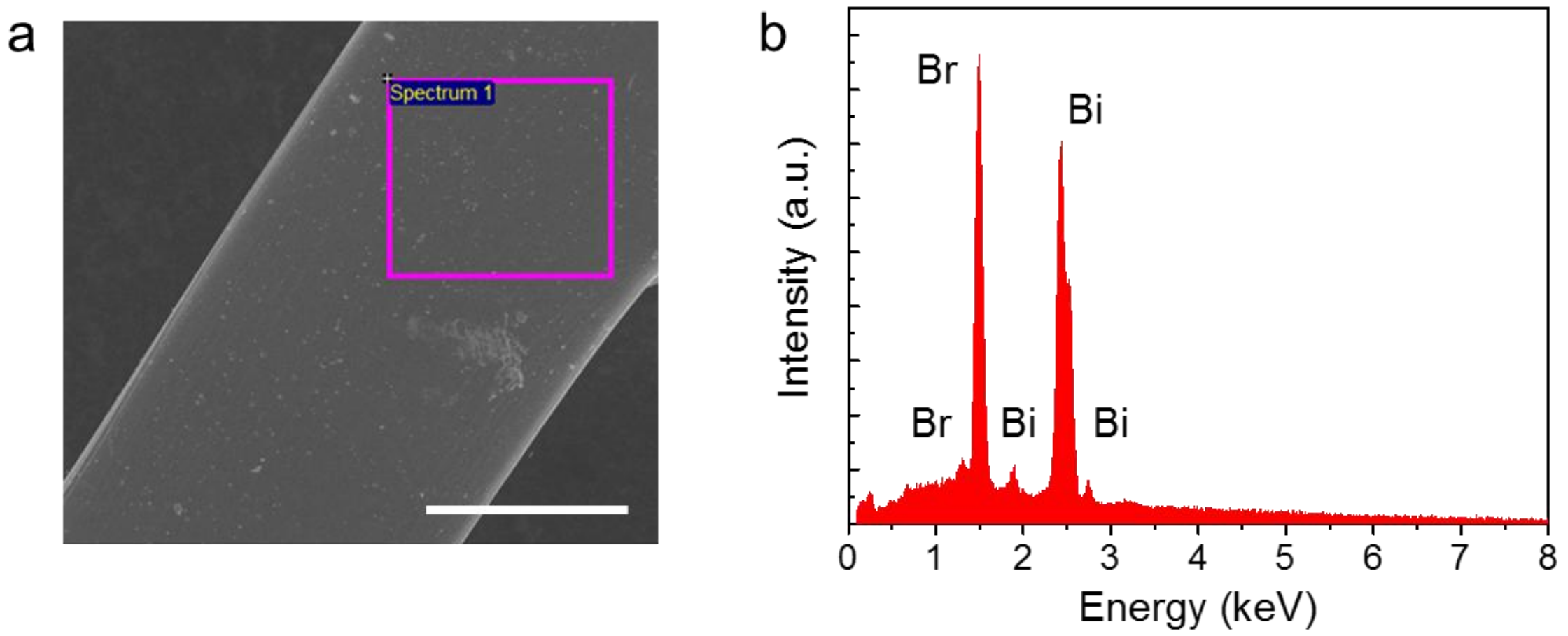

**Supplementary Figure 3 | a, b,** SEM image and EDS analyses of α-$Bi_4Br_4$ crystals. Scale bar, 100 μm.

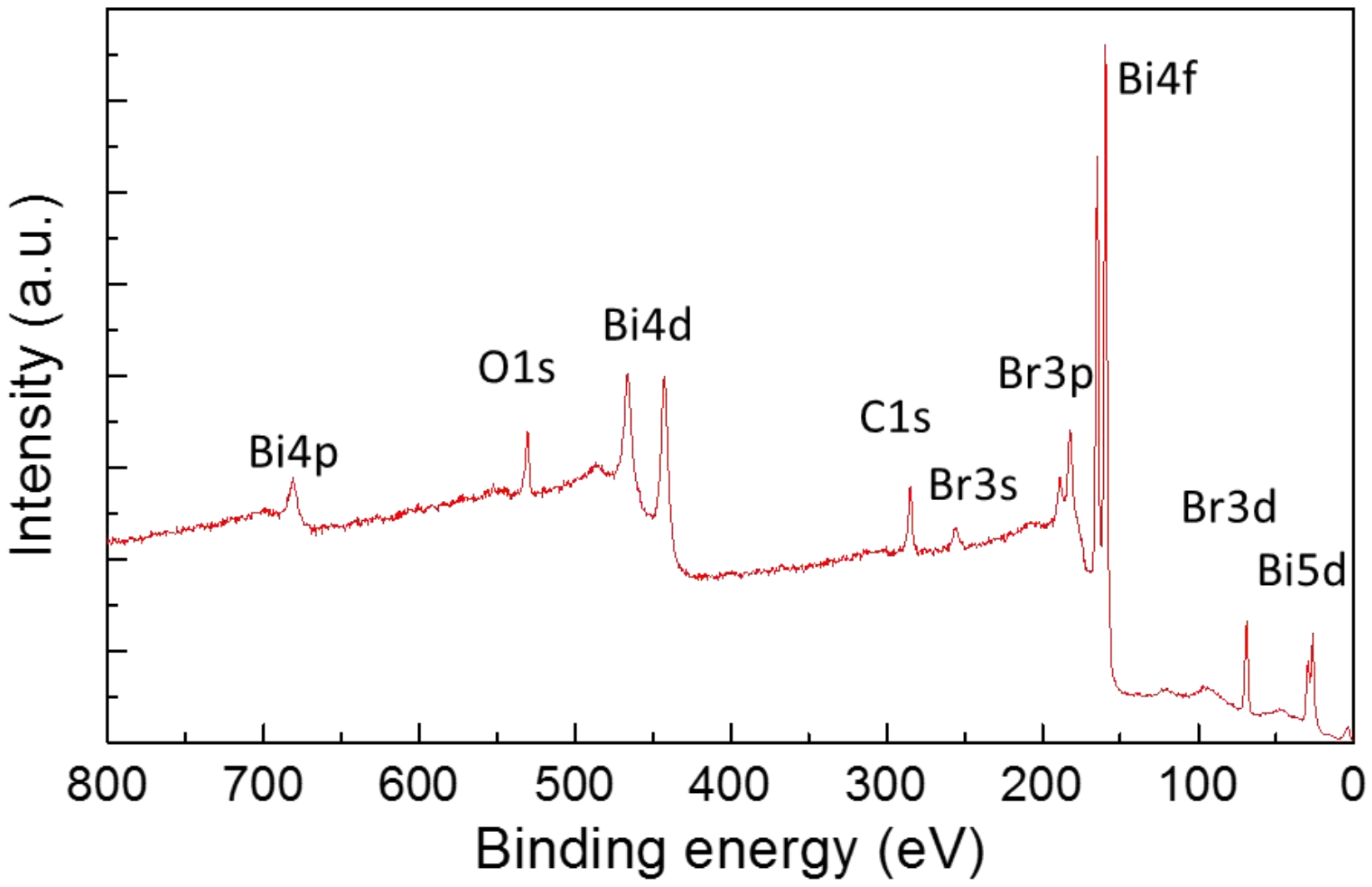

**Supplementary Figure 4 |** XPS spectra of α-$Bi_4Br_4$ crystals.

The α-$Bi_4Br_4$ belts were obtained by mechanical exfoliation and transferred onto $CaF_2$ substrates. In the first step, the adhesive tape was attached to a selected high-quality α-$Bi_4Br_4$ single crystal. After exfoliation, the tape loaded with α-$Bi_4Br_4$ belts was brought in contact with a $CaF_2$ substrate. After holding for several seconds, the tape was removed and various α-$Bi_4Br_4$ belts were transferred onto the $CaF_2$ substrate. The thickness and step profiles were determined using profilemeter. Due to the weak vdW-type interlayer and interchain coupling in α-$Bi_4Br_4$, it is easy to obtain α-$Bi_4Br_4$ belts with (001)-oriented flat tops along the ab plane and straight edges along the 1D chain axis. These long and straight edges without dangling bonds are predicted to host topologically edge states, according to the theory discussed in the main text.

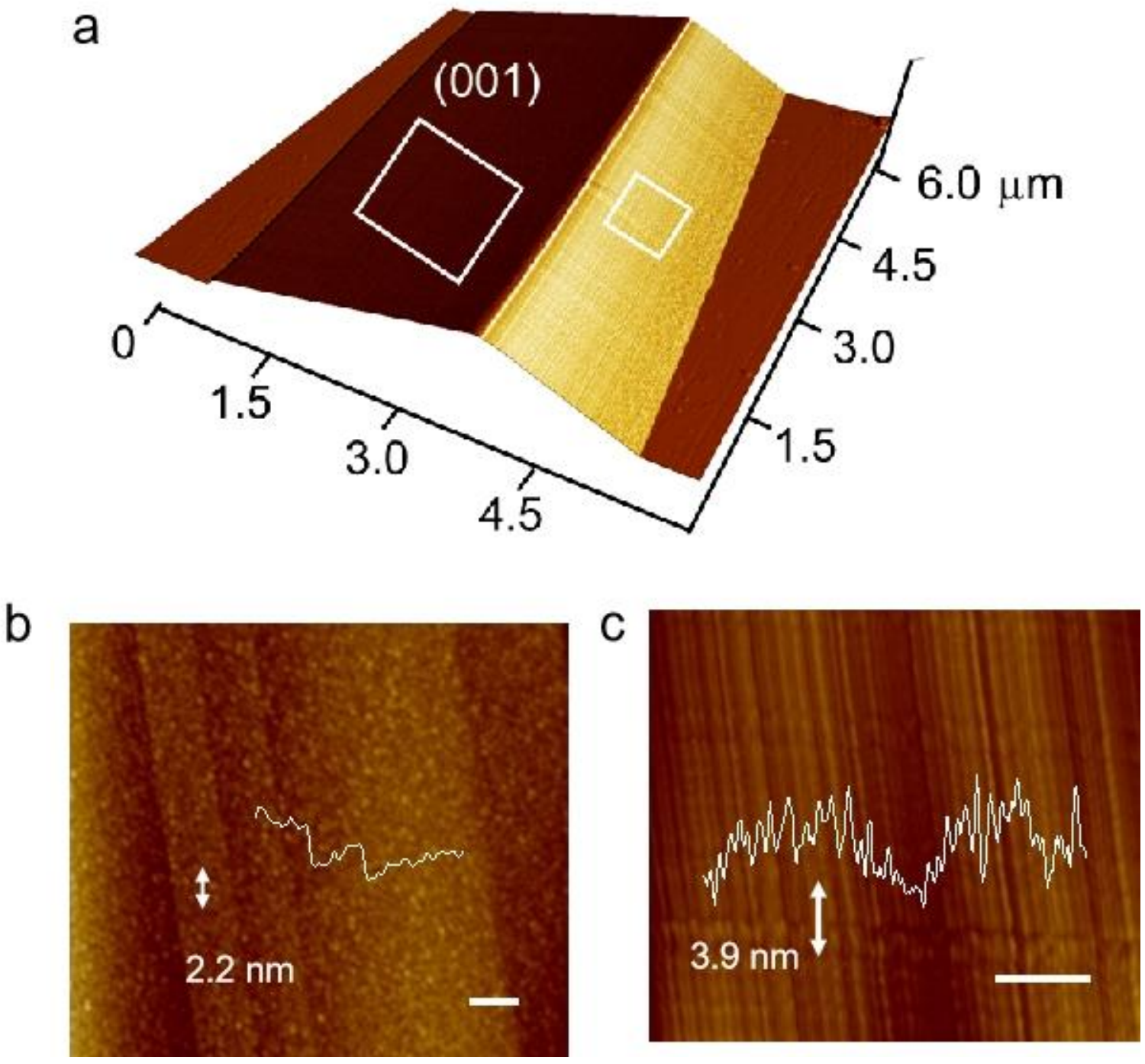

**Supplementary Figure 5 | a,** AFM 3D image of an α-$Bi_4Br_4$ belt. **b, c,** Zoom-in AFM images acquired at the top and edge of the belt, respectively. Scale bar, 100 nm.

To directly reveal the structure of the edges of the belts, we performed atomic force microscopy measurements. Supplementary Fig. 5 depicts a large-scale 3D atomic force microscopy image of an as-prepared α-$Bi_4Br_4$ belt. The zoom-in AFM images and line profile analysis indeed reveal that the (001)-oriented top of the belt is very flat with a few atomic-height steps, while a bunch of straight edges can be observed at the edges of the belt. These long and straight edges without dangling bonds are predicted to host topologically originated edge states, according to the theory in the main text. Therefore, although our optical measurements cannot provide a spatial resolution of nanometer scale for the edge states at each individual step edge, we expect that the edges of the α-$Bi_4Br_4$ belt exhibit a much stronger infrared absorption than the top surface, as their step density is much higher than that of the top surface.

**Supplementary Note 2: Infrared absorption measurements**

The optical image and infrared absorption spectra were acquired using a Fourier transform infrared microscopy (Bruker HYPERION 3000) with a pair of 15× reflective microscope objectives in transmission mode. The detection spectral range extended from 4000 $cm^{-1}$ to 900 $cm^{-1}$ (photon energy of 0.11 eV-0.5 eV). As shown in Supplementary Fig. 6, the incident light has a tilting angle of 12-23.6° away from the normal of the belt top and can illuminate both the top and the side of the belts, which enlarge the observed region of the belt edge. A spatial resolution of the infrared absorption maps is 7-8 μm in the wavenumber region of 800 $cm^{-1}$ ~ 4000 $cm^{-1}$ (photon energy of 0.11 eV - 0.5 eV). In polarization-dependent infrared absorption measurements, we acquired polarized light with a $CaF_2$ holographic wire grid polarizer in front of the condenser. The infrared absorption integration image was exported from OPUS software, which was used to control the FTIR microscopy and analyze the absorption spectrum.

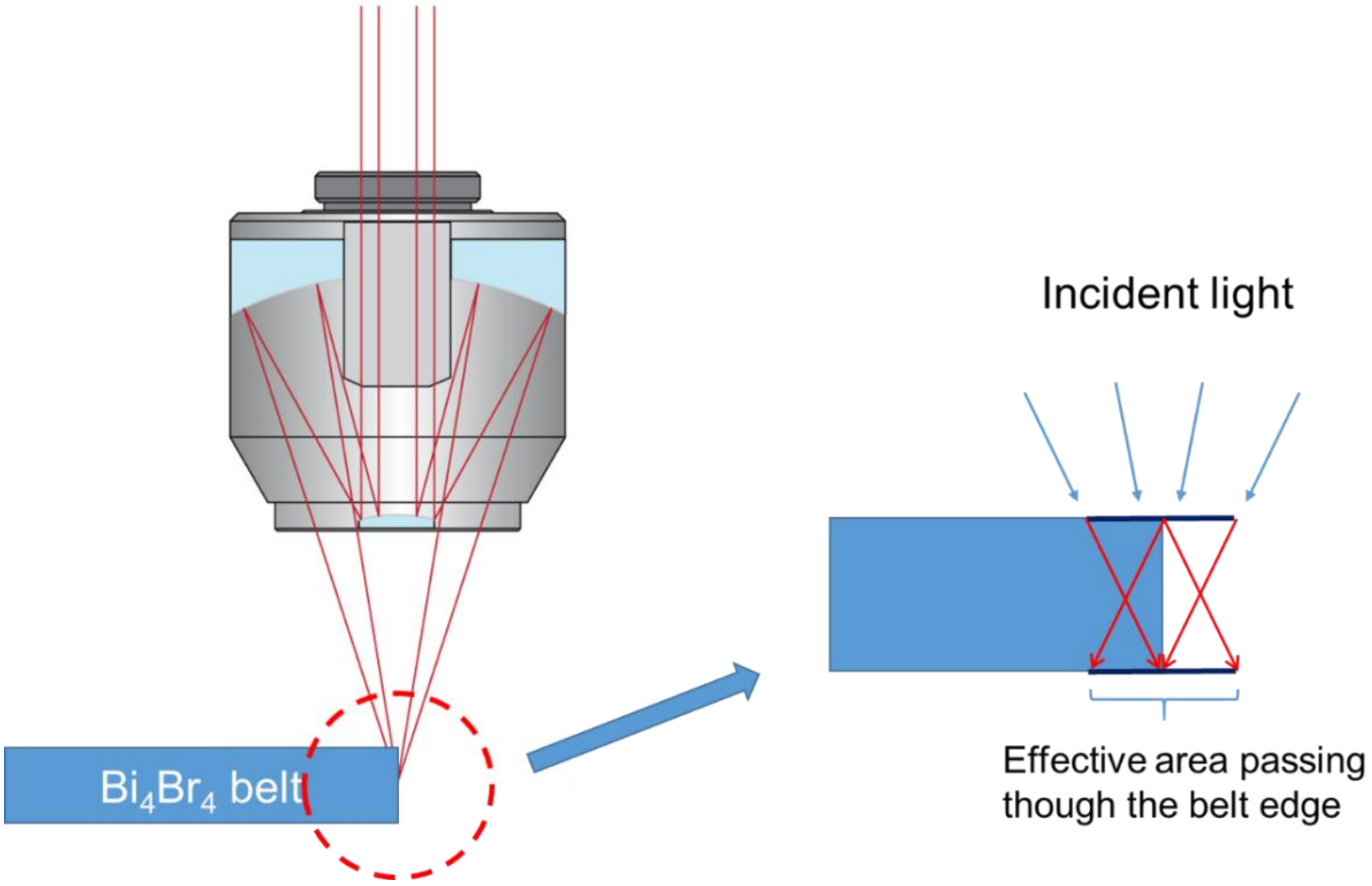

**Supplemental Figure 6 | Schematic of the objective for infrared spectroscopy measurements.**

For the absorption spectra (red curve) acquired at the centre of the belt, the obvious absorption slope in the region of 0.22-0.24 eV can be fitted by the classical Tauc plot for optical absorption in a semiconductor[1], and a band gap of 0.22 eV is estimated for bulk α-$Bi_4Br_4$ as shown in Supplementary Fig. 7.

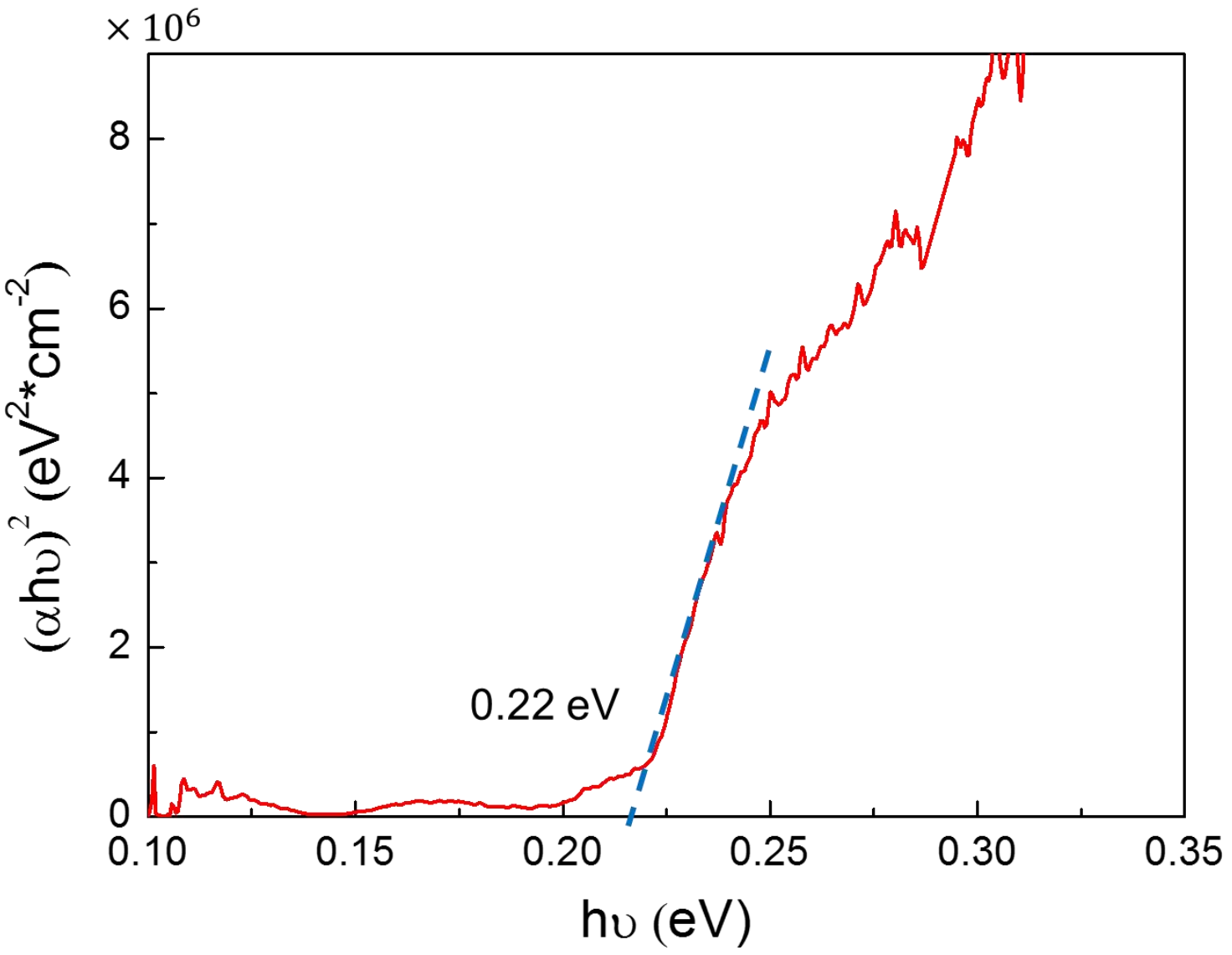

**Supplementary Figure 7 | Plots of $(\alpha h\upsilon)^2$ vs. hυ.** **α, h** and **υ** are absorption coefficient, Planck constant and photon frequency, respectively. An optical band gap of ~ 0.22 eV is derived.

As shown in Supplementary Fig. 8 shows a series of background-subtracted infrared spectra collected along a line perpendicularly across the α-$Bi_4Br_4$ belt. In Fig. S8a, The yellow region is the α-$Bi_4Br_4$ belt, while the blue region is the $CaF_2$ substrate. With the incident light moving from the center to the side, less light illuminates the α-$Bi_4Br_4$ belt, resulting in a reduction of its infrared absorption. Indeed, in the large photon energy region, the absorption is progressively reduced from the center to the side (from the red line to the blue line). However, the infrared absorption in the small photon energy region (< 0.2 eV) is surprisingly stronger than that from the center.

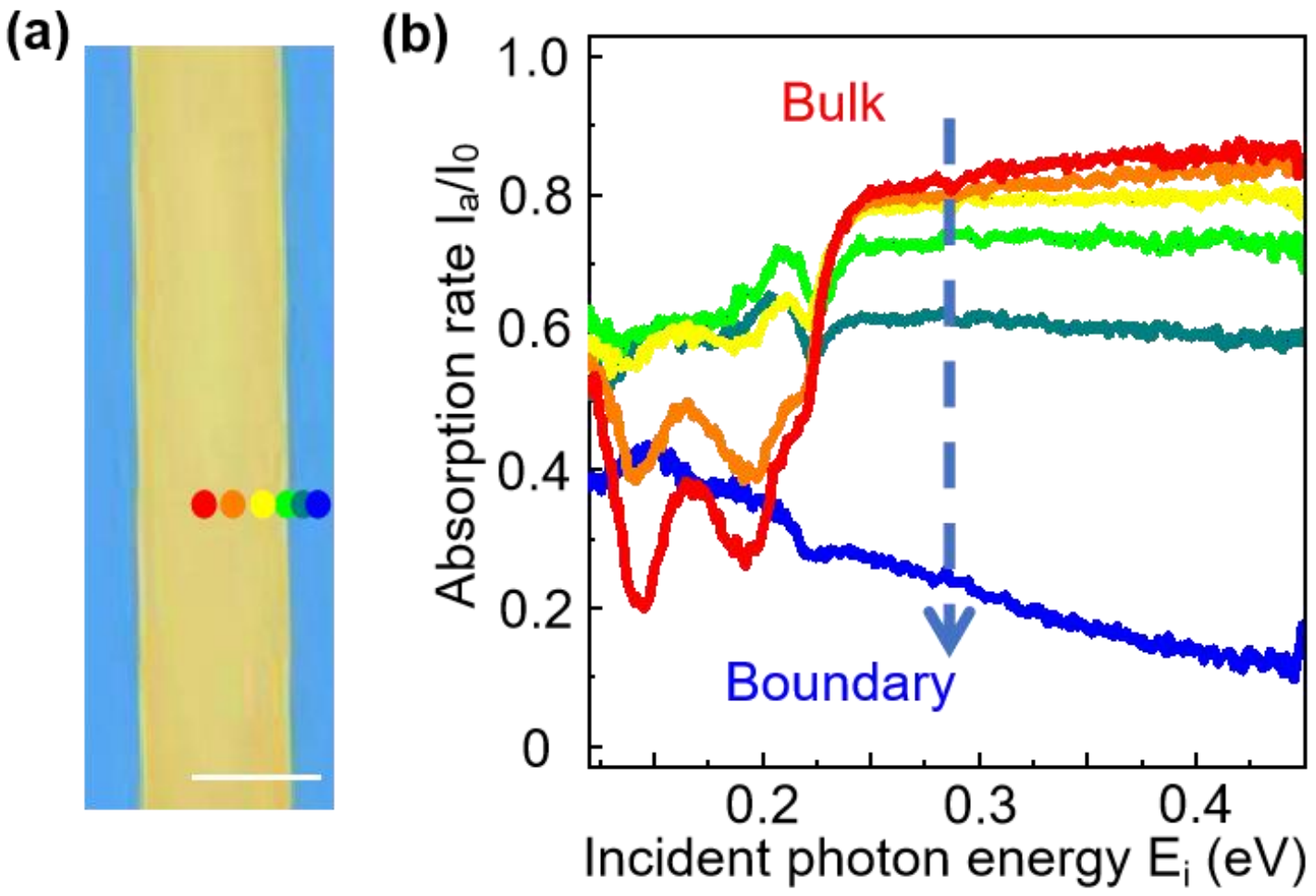

**Supplementary Figure 8 | Line scan analysis of infrared spectra of α-$Bi_4Br_4$ belt.** **a,** OM image of the belt. Scale bar, 40 μm. **b,** Infrared absorption spectra acquired from the belt centre to the belt side. The positions where the spectra were acquired are indicated by the dots in **a**.

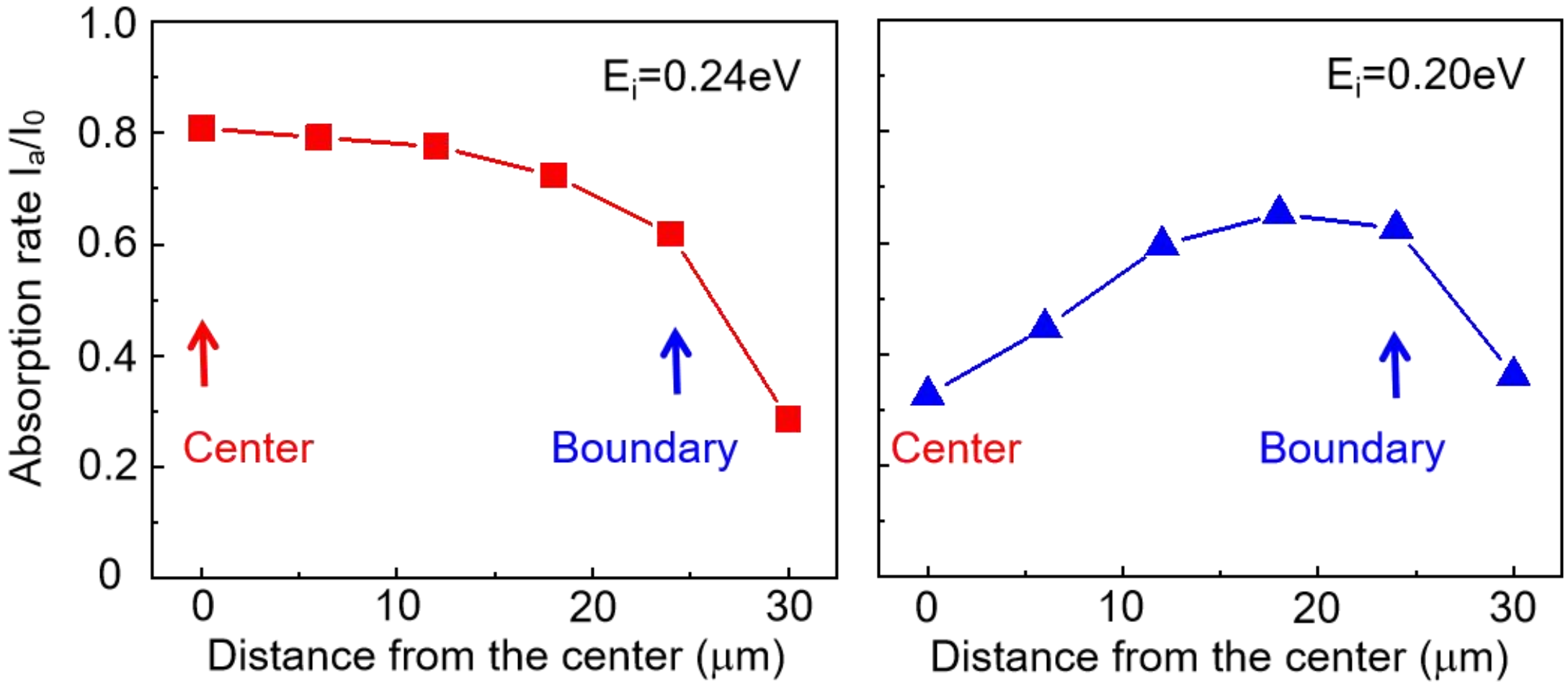

**Supplementary Figure 9 |** Absorption rate VS distance from the center of α-$Bi_4Br_4$ belt with incident photon energy of 0.24 eV and 0.20 eV.

**Supplementary Note 3: infrared-pump infrared-probe analyses**

Ultrafast infrared-pump infrared-probe measurements were performed with a femtosecond Ti:Sapphire regenerative amplifier (Spitfire Ace, 3.5W, Spectra Physics). The amplifier generated 35 fs pulses at a repetition rate of 1 kHz and the central wavelength of the pulse is 800 nm. The laser beam was split into two parts, one was used to pump an optical parametric amplifier (TOPAS Prime, Spectra Physics) to generate infrared pump beam (0.5 eV) and the other was used to generate ultra-broadband super-continuum pulse (photon energy range of 0.12 eV-0.45 eV). The key techniques were to obtain a wide-energy-range probe light by forming a plasma filament (0.12-0.45 eV). A 0.17 eV pump pulse was obtained with difference frequency generation extension after optical parametric amplifier. Probe beam was focused to the sample with a 15× reflective microscope objective (LMM-15X-P01, Thorlabs), diameter of focal spot is about 20 μm at 7 μm wavelength. The pump beam is focused to the sample with a $CaF_2$ planoconvex lens (focal length = 500 mm), noncollinear with probe beam. The diameter of pump focal spot is about 600 μm at 7 μm wavelength and its overlap with probe beam is optimized based on the measurement signal intensity. A DC servo motor actuator (Z812B, Thorlabs) was used to drive a translation stage which provided high-resolution linear motion of the α-$Bi_4Br_4$ belt respect to the pump and probe beam spot. After transmitted from the sample area, the probe light was detected by a grating spectrograph with liquid-nitrogen-cooled 64×2 mercury–cadmium–telluride array detector. The setup was powerful to investigate the carrier dynamics in the energy range around the band gap of $Bi_4Br_4$.

In Fig. 4c of the main text, the signal from the boundary is performed as a saturation behavior under increasing pump fluences. The fitted saturation fluence is ~35 μJ/cm$^2$. Considering the incident photon energy of 0.17 eV, the photon density is ~$1.29\times10^{15}$ cm$^{-2}$ in each laser pulse. Since the incident angle is around 45° between incident light and belt surface, a simple geometry calculation indicates that ~$1.18\times10^{7}$ photons can pass through the belt side per micrometer along the *b* axis. The illustrated belt side is composed of ~$1.57\times10^{6}$ unit cells per micrometer along the *b* axis, which contains ~$6.28\times10^{6}$ edge states. Thus, the incident light with fluences of ~35 μJ/cm$^2$ per pulse is enough to excite most carriers in edge states and cause a saturation behavior.

Table S1 Carrier lifetime of topological materials

| Materials | Method | Pump photon energy | Life time | Reference |
|---|---|---|---|---|
| $Bi_2Se_3$ | spatially resolved pump-probe technique | 1.55 eV pump | 3-10 ps at 300K | 2 |
| $Bi_2Se_3$ | rotational anisotropy time-resolved SHG | 1.56 eV pump | 20 ps at 300 K | 3 |
| $Bi_2Se_3$ | time and angle-resolved photoelectron spectroscopy | 1.55 eV pump | 15 ps at 15 K | 4 |
| $Bi_2Se_3$ | Transient reflectivity | 1.51 eV pump | 5-220 ps at 300K | 5 |
| $Bi_2Se_3$ | Transient reflectivity | 1.51 eV pump | 2 ps at 300K | 6 |
| $Bi_2Se_3$ | Transient mid-IR reflectance response | 1.5eV pump 0.3-1.2 eV probe | 150 ps at 10K 50 ps at 300K | 7 |
| $Bi_2Te_3$ | time and angle-resolved photoelectron spectroscopy | 1.58eV pump | 10 ps at 20K | 8 |
| $Bi_2Te_3$ | time and angle-resolved photoelectron spectroscopy | 1.5 eV pump | 45 ps at 130 K | 9 |
| $Sb_2Te_3$ | two-photon photoemission | 2.58 eV pump | 2-4 ps at 300K | 10 |
| $Bi_2Te_2Se$ | time and angle-resolved photoelectron spectroscopy | 1.47 eV pump | 4 μs * | 11 |
| $Bi_2Te_2Se$ | time and angle-resolved photoelectron spectroscopy | 1.48 eV pump | 15 ps at 7 K | 12 |
| $Bi_2Te_2Se$ | optical pump and probe measurements | 0.17 eV pump | 30 ps at 300K | 13 |
| $Bi_{1.5}Sb_{0.5}Te_{1.7}Se_{1.3}$ | mid-IR pump-probe reflectivity measurements | 0.3-0.66 eV pump | 10 ps at 300K | 14 |
| $Bi_xSb_{2-x}Te_ySe_{3-y}$ | time and angle-resolved photoelectron spectroscopy | 1.2 eV pump | 2-8 ps at 300 K | 15 |
| $Bi_{2-x}Sb_xSe_3$ | time-resolved transient reflectivity | 1.55 eV pump | 3.3 ns at 7 K | 16 |

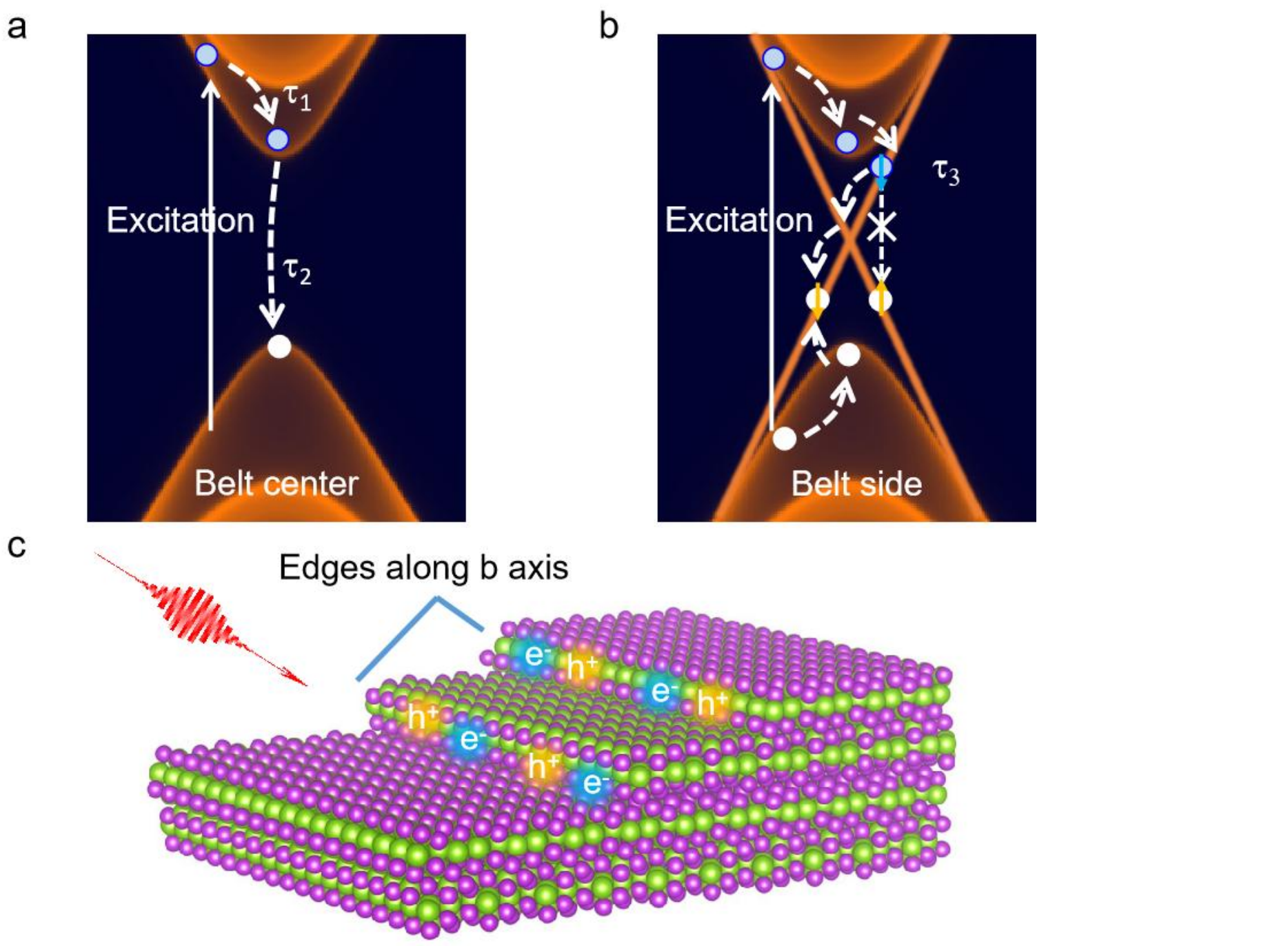

**Supplementary Figure 10 | Illustration of carrier dynamics of the edge states of α-$Bi_4Br_4$. a,** Schematic representation of the ultrafast excitation and subsequent relaxation processes of photoexcited carriers in the conduction band and the valence band. (1) Photoexcitation of electron-hole pairs; (2) hot carrier cooling due to electron-electron scattering or electron-phonon scattering; (3) recombination of electrons and holes in conduction band and valence band, respectively; **b,** Schematic representation of relaxation processes of photoexcited carriers in the edge states. (1) Direct recombination of electrons

and holes is prohibited due to their opposite spin polarization; (2) the electron-phonon scattering with the same spin polarization is suppressed by the very limited scattering phase space; **c,** Illustration of the inter-layer scattering of carriers localized at different layers. The edges of different layers are not strictly aligned, forming step edges with decoupled states, which could significantly suppress the recombination of electrons and holes by the inter-layer scattering.

## Supplementary Note 4: theoretical calculations

To calculate the electronic structure, the first-principles calculations were performed by using Vienna *ab initio* simulation package (VASP) based on the density function theory with Perdew-Burke-Ernzerhof (PBE) parameterization of generalized gradient approximation (GGA)[17-19] and the Heyd–Scuseria–Ernzerhof hybrid (HSE06) functional[20,21]. The energy cutoff of the plane wave basis was set as 300 eV, and the Brillouin zone of three- (two-) dimensional $Bi_4Br_4$ was sampled by a 9 × 9 × 6 (6 × 6 × 1) k-mesh. The ionic positions were fully optimized with the vdW correction until the force on each atom was less than 0.01 eV/Å while the lattice parameters were fixed to the experimental value. The surface states were calculated based on maximally localized Wannier functions[22] and surface Green function method[23].

To calculate the optical absorption coefficients, a slab consisting of 20 unit cells (25 nm width along the *a*-direction) was used to model the α-$Bi_4Br_4$ belt, and its Hamiltonian was constructed with 1920 Wannier basis. Therefore, dielectric constants and infrared absorption coefficients can be calculated by the formula as follows[24,25]:

$$\operatorname{Im}\varepsilon_{\alpha\beta}(\omega)=\frac{4\pi^2e^2\hbar^4}{\Omega\omega^2m_e^2}\sum_{v,c,k}2w_k\delta(E_{ck}-E_{vk}-\omega)\langle ck|i\nabla_\alpha-k_\alpha|vk\rangle\langle ck|i\nabla_\beta-k_\beta|vk\rangle^* \quad (1)$$

$$\operatorname{Re}\varepsilon_{\alpha\beta}(\omega)=1+\frac{2}{\pi}P\int_0^\infty\frac{\operatorname{Im}\varepsilon_{\alpha\beta}(\omega')\omega'}{\omega'^2-\omega^2}d\omega' \quad (2)$$

$$n(\omega)=\sqrt{\frac{1}{2}\left[\operatorname{Re}\varepsilon(\omega)+\sqrt{\left(\operatorname{Re}\varepsilon(\omega)\right)^2+\left(\operatorname{Im}\varepsilon(\omega)\right)^2}\right]} \quad (3)$$

$$A(\omega)=\frac{\omega\operatorname{Im}\varepsilon(\omega)}{n(\omega)c} \quad (4)$$

Here, Ω is the volume of the primitive cell, $w_k$ is the k-point weights, $m_e$ is the mass of electron, and *ck* (*vk*) represents the electronic state of conduction (valence) band at *k* point. According to equation (1), we can first get the imaginary part of dielectric constants. Then the real part of dielectric constants can be calculated using Kramers-Kronig transformation (equation (2)). Furthermore, refractivity $n(\omega)$ and absorption spectra $A(\omega)$ can be obtained easily from equation (3) and (4), respectively.

## Supplementary references